\documentclass[aps,prd,twocolumn,superscriptaddress,showpacs]{revtex4}

\usepackage[dvips]{graphicx}
\usepackage{ulem}
\usepackage[usenames]{color}
\usepackage{amsmath}
\usepackage{float}

\begin{document}


\title{Hadronic decay properties of newly observed $\Omega_c$ baryons}


\author{Ze Zhao}
\affiliation{Department of Physics, Shanghai University, Shanghai 200444, China}

\author{Dan-Dan Ye}
\affiliation{Department of Physics, Shanghai University, Shanghai 200444, China}
\affiliation{College of Mathematics, Physics and Information Engineering, Jiaxing University, Jiaxing 314001, China}

\author{Ailin Zhang}
\email{zhangal@staff.shu.edu.cn}
\affiliation{Department of Physics, Shanghai University, Shanghai 200444, China}

\begin{abstract}
Hadronic decay widths of the newly observed charmed strange baryons, $\Omega_c(3000)^0$, $\Omega_c(3050)^0$, $\Omega_c(3066)^0$, $\Omega_c(3090)^0$ and $\Omega_c(3119)^0$ have been calculated in a $^3P_0$ model. Our results indicate that $\Omega_c(3066)^0$ and $\Omega_c(3090)^0$ can be interpreted as the $1P-$wave $\Omega_{c2}(\frac{3}{2}^-)$ or $\Omega_{c2}(\frac{5}{2}^-)$. Though the measured masses of $\Omega_c(3000)^0$, $\Omega_c(3050)^0$ and $\Omega_c(3119)^0$ are lower than existed theoretical predictions of $1D-$wave $\Omega_c$, the hadronic decay features of these $\Omega_c$ favor assignments of the $1D-$wave states. $\Omega_c(3000)^0$ is possibly $\Omega_{c1}(\frac{1}{2}^+)$ or $\Omega_{c1}(\frac{3}{2}^+)$, $\Omega_c(3050)^0$ is possibly $\hat\Omega_{c3}(\frac{5}{2}^+)$ or $\hat\Omega_{c3}(\frac{7}{2}^+)$, and $\Omega_c(3119)^0$ is possibly $\hat\Omega_{c3}(\frac{5}{2}^+)$, $\hat\Omega_{c3}(\frac{7}{2}^+)$, $\Omega_{c3}(\frac{5}{2}^+)$ or $\Omega_{c3}(\frac{7}{2}^+)$. The predicted total decay widths in these assignments are consistent with experiment.

\end{abstract}


\maketitle

\section{Introduction \label{sec:introduction}}
Very recently, five narrow excited $\Omega_c^0$ baryons were reported by the LHCb collaboration~\cite{LHCb1}. Their masses and decay widths were measured in $\Omega_c^0 \to \Xi_c^+K^-$ as follows,
\begin{eqnarray*}
 \Omega_c(3000)^0:  M & = 3000.4 \pm 0.2 \pm 0.1 ^{+0.3}_{-0.5}~{\rm MeV} \\  \Gamma  &= 4.5 \pm 0.6 \pm 0.3~{\rm MeV}
\end{eqnarray*}
    \begin{align*}
     \Omega_c(3050)^0 : M & = 3050.2 \pm 0.1 \pm 0.1 ^{+0.3}_{-0.5}~{\rm MeV} \\  \Gamma & = 0.8 \pm 0.2 \pm 0.1~{\rm MeV}
    \end{align*}
\begin{align*}
 \Omega_c(3066)^0 : M & = 3065.6 \pm 0.1 \pm 0.3 ^{+0.3}_{-0.5}~{\rm MeV} \\ \Gamma & = 3.5 \pm 0.4 \pm 0.2~{\rm MeV}
\end{align*}
    \begin{align*}
     \Omega_c(3090)^0 :M & = 3090.2 \pm 0.3 \pm 0.5 ^{+0.3}_{-0.5}~{\rm MeV} \\ \Gamma & = 8.7 \pm 1.0 \pm 0.8~{\rm MeV}
    \end{align*}
\begin{align*}
 \Omega_c(3119)^0 :M & = 3119.1 \pm 0.3 \pm 0.9 ^{+0.3}_{-0.5}~{\rm MeV} \\ \Gamma & = 1.1 \pm 0.8 \pm 0.4~{\rm MeV}
\end{align*}
The $J^P$ quantum numbers of these five $\Omega_c^0$ baryons have not been measured.

According to Particle Data Group~\cite{pdg2016}, two $\Omega_c$ baryons have been observed: the ground state $\Omega_c^0$ and $\Omega_c(2770)^0$ (also known as $\Omega_c^\ast$) with $J^P=\frac{1}{2}^+$ and $\frac{3}{2}^+$, respectively. No other $\Omega_c$ baryon has been reported before the LHCb experiment.

Theoretical predictions of the spectra of orbitally excited $\Omega_c$ baryons have been conducted in many different models~\cite{Capstick:2809 (1986),Ebert2011, Roberts2008, Gar2007, Yoshida, Shah}. Relevant references can be found in reviews~\cite{Capstick:S241 (2000),klempt,roberts} and references therein. From the spectra, these newly observed baryons are most possibly the orbitally excited $1P$ or the radially excited $2S$ states. Theoretical predicted masses of the orbitally excited $1D$ $\Omega_c$ baryons are $100-200$ MeV higher than the observed ones.

Since the report of LHCb, the spectra of orbitally excited $\Omega^0_c$ baryons have been explored in Refs.~\cite{Agaev2017,Marek2017,Padm,wwang}. The hadronic decays of these baryons have also been studied through QCD sum rules, chiral quark model or other models in Refs.~\cite{CHX2017,Marek2017,Zhong2017,CHY2017,wwang}. In these explorations, the observed $\Omega^0_c$ baryons were interpreted as the $1P$ or $2S$ $\Omega_c$, while the $1D$ possibility has not been explored for their lower masses. However, the mass prediction may be largely different in different models. As indicated in Ref.~\cite{chen}, the mass prediction of the excited $\Lambda_c$ and $\Xi_c$ baryons in Ref.~\cite{Capstick:2809 (1986)} were really largely different from the results in Ref.~\cite{Ebert2011}. The study of $1D$ possibility of these $\Omega^0_c$ through their hadronic decays is necessary.

For the understanding of these excited baryons, it is useful to explore their hadronic decays in different model for a cross-check. $^3P_0$ model is a phenomenological method to study the OZI-allowed hadronic decays of hadrons. In addition to mesons, it is employed successfully to explain the hadronic decays of baryons~\cite{yaouanc1,Capstick:2809 (1986),Roberts:171 (1992),Capstick:1994 (1993),Capstick:4507 (1994),Capstick:S241 (2000),Chong:094017 (2007)}. In this paper, $^3P_0$ model will be employed to study the hadronic decays of $\Omega_c(3000)^0$, $\Omega_c(3050)^0$, $\Omega_c(3066)^0$, $\Omega_c(3090)^0$ and $\Omega_c(3119)^0$. In particular, the $1D$ possibility of these $\Omega_c^0$ will be examined in detail.

This work is organized as follows. In Sec.II, we give a brief review of the $^3P_0$ model. In Sec.III, we present our numerical results. We give our conclusions and discussions in Sec.IV.

\section{Baryon decay in the $^3P_0$ model \label{Sec: $^3P_0$ model}}
 Proposed by Micu\cite{micu1969} and later developed by Yaouanc et al~\cite{yaouanc1,yaouanc2,yaouanc3}, $^3P_0$ model is also known as a Quark Pair Creation (QPC) model. It assumes that a pair of quark $q\bar{q}$ is created from the vacuum with $J^{PC}=0^{++}(^{2S+1}L_J=$ $^3P_0)$, and then the created quarks regroup with the quarks from the initial hadron A to form two daughter hadrons B and C. The process of a baryon decay is shown in Fig.~I.
\begin{figure}\label{figure1}
\begin{center}
\includegraphics[height=2.8cm,angle=0,width=6cm]{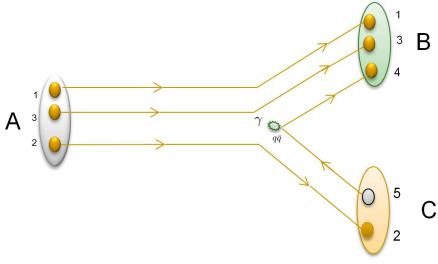}
\caption{Baryon decay process of $A\to B+C$ in the $^3P_0$ model.}
\end{center}
\end{figure}

In the $^3P_0$ model, the hadronic decay width $\Gamma$ of a process $A \to B + C$ is as follows~\cite{yaouanc3},
\begin{eqnarray}
\Gamma  = \pi ^2 \frac{|\vec{p}|}{m_A^2} \frac{1}{2J_A+1}\sum_{M_{J_A}M_{J_B}M_{J_C}} |{\mathcal{M}^{M_{J_A}M_{J_B}M_{J_C}}}|^2,
\end{eqnarray}
where $m_A$ and $J_A$ are the mass and total angular momentum of the initial baryon A, respectively. $m_B$ and $m_C$ are the masses of the final hadrons. $\mathcal{M}^{M_{J_A}M_{J_B}M_{J_C}}$ is the helicity amplitude, which reads~\cite{Chong:094017 (2007)}
\begin{flalign}
 &\mathcal{M}^{M_{J_A } M_{J_B } M_{J_C }}\nonumber \\
 &=-2\gamma\sqrt {8E_A E_B E_C }  \sum_{M_{\rho_A}}\sum_{M_{L_A}}\sum_{M_{\rho_B}}\sum_{M_{L_B}} \sum_{M_{S_1},M_{S_3},M_{S_4},m}  \nonumber\\
 &\langle {J_{l_A} M_{J_{l_A} } S_3 M_{S_3 } }| {J_A M_{J_A } }\rangle \langle {L_{\rho_A} M_{L_{\rho_A} } L_{\lambda_A} M_{L_{\lambda_A} } }| {L_A M_{L_A } }\rangle \nonumber \\
 &\langle L_A M_{L_A } S_{12} M_{S_{12} }|J_{L_A} M_{J_{L_A} } \rangle \langle S_1 M_{S_1 } S_2 M_{S_2 }|S_{12} M_{S_{12} }\rangle \nonumber \\
 &\langle {J_{l_B} M_{J_{l_B} } S_3 M_{S_3 } }| {J_B M_{J_B } }\rangle \langle {L_{\rho_B} M_{L_{\rho_B} } L_{\lambda_B} M_{L_{\lambda_B} } }| {L_B M_{L_B } }\rangle \nonumber \\
 &\langle L_B M_{L_B } S_{14} M_{S_{14} }|J_{14} M_{J_{14} } \rangle \langle S_1 M_{S_1 } S_4 M_{S_4 }|S_{14} M_{S_{14} }\rangle \nonumber \\
 &\langle {1m;1 - m}|{00} \rangle \langle S_4 M_{S_4 } S_5 M_{S_5 }|1 -m \rangle \nonumber \\
 &\langle L_C M_{L_C } S_C M_{S_C}|J_C M_{J_C} \rangle \langle S_2 M_{S_2 } S_5 M_{S_5 }|S_C M_{S_C} \rangle \nonumber \\
&\times\langle\varphi _B^{1,4,3} \varphi _C^{2,5}|\varphi _A^{1,2,3}\varphi _0^{4,5} \rangle \times I_{M_{L_B } ,M_{L_C } }^{M_{L_A },m} (\vec{p}).
\end{flalign}

The space integral $I_{M_{L_B } ,M_{L_C } }^{M_{L_A } ,m} (\vec{p})$ follows as~\cite{114020}
\begin{flalign}
&\delta^3(B+C) I_{M_{L_B } ,M_{L_C } }^{M_{L_A } ,m} (\vec{p}) \nonumber \\
&= \int d \vec{p}_1 d \vec{p}_2 d \vec{p}_3 d \vec{p}_4 d \vec{p}_5 \nonumber \\
&\times\delta ^3 (\vec{p}_1 + \vec{p}_2 + \vec{p}_3 -\vec{p}_A)\delta ^3 (\vec{p}_4+ \vec{p}_5)\nonumber \\
&\times \delta ^3 (\vec{p}_1 + \vec{p}_4 + \vec{p}_3 -\vec{p}_B )\delta ^3 (\vec{p}_2 + \vec{p}_5 -\vec{p}_C) \nonumber \\
& \times\Psi _{B}^* (\vec{p}_1, \vec{p}_4,\vec{p}_3)\Psi _{C}^* (\vec{p}_2 ,\vec{p}_5) \nonumber \\
& \times \Psi _{A} (\vec{p}_1 ,\vec{p}_2 ,\vec{p}_3)y _{1m}\left(\frac{\vec{p_4}-\vec{p}_5}{2}\right).
\end{flalign}

Simple harmonic oscillator (SHO) wave functions are employed as the baryon wave functions\cite{Capstick:2809 (1986),Capstick:1994 (1993),Capstick:4507 (1994)}. For practical calculations such as the flavor matrix $\langle \varphi_B^{1,4,3} \varphi_C^{2,5}|\varphi_A^{1,2,3}\varphi_0^{4,5} \rangle$ and the space integrals in the model, more details were presented in Refs.\cite{Capstick:1994 (1993),yaouanc3,Chong:094017 (2007), 114020}.

Two parameters, the $\beta$ in SHO and the quark pair creation strength $\gamma$, are adopted as those in Refs.~\cite{Chong:094017 (2007),Blundell:3700 (1996), 114020}. $\gamma=13.4$, and $\beta=476$ MeV for meson $K$. For baryons, a universal value $\beta=600$ MeV is employed. The masses of $\Omega_c(3000)^0$, $\Omega_c(3050)^0$, $\Omega_c(3066)^0$, $\Omega_c(3090)^0$ and $\Omega_c(3119)^0$ are taken as $3000.4$ MeV, $3050.2$ MeV, $3065.6$ MeV, $3090.2$ MeV and $3119.1$ MeV, respectively. Masses of K mesons and $\Xi_c$ baryons are taken from PDG~\cite{pdg2016}.

The quantum numbers of the P-wave and D-wave baryons are complicated~\cite{CHX2015,CHX2016}, and we follow the notations in Ref.~\cite{Chong:094017 (2007)}. The quantum numbers involved in the calculations are listed in Table~\ref{quantum numbers}. In the table, $L_\rho$ denotes the orbital angular momentum between the two light quarks, $L_\lambda$ denotes the orbital angular momentum between the charm quark and the two light quark system, $L$ is the total orbital angular momentum of $L_\rho$ and $L_\lambda$. $S_\rho$ denotes the total spin of the two light quarks, $J_l$ is total angular momentum of $L$ and $S_\rho$. $J$ is the total angular momentum of the baryons. The hat and the check are also used to denote the assignments with $L_\rho=2$ and $L_\rho=1$, respectively. The superscript $L$ is adopted to denote the different total orbital angular momentum in $\check\Omega_{cJ_l}^{\ L}$.
\begin{table}[t]
\caption{Quantum numbers of initial baryons}
\begin{tabular}{p{0.0cm} p{2.5cm}*{6}{p{0.8cm}}}
   \hline\hline
             & Assignments                                        & $J$                         & $J_l$ & $L_\rho$ & $L_\lambda$ & $L$  & $S_\rho$ \\
   \hline
\label{P01  }&$\Omega_{c0}(\frac{1}{2}^-)$                        & $\frac{1}{2}$               &  0    &  0       &   1         &  1   &  1       \\
\label{P0203}&$\Omega_{c1}(\frac{1}{2}^-, \frac{3}{2}^-)$         & $\frac{1}{2}$,$\frac{3}{2}$ &  1    &  0       &   1         &  1   &  1       \\
\label{P0405}&$\Omega_{c2}(\frac{3}{2}^-, \frac{5}{2}^-)$         & $\frac{3}{2}$,$\frac{5}{2}$ &  2    &  0       &   1         &  1   &  1       \\
\label{P0607}&$\tilde\Omega_{c1}(\frac{1}{2}^-, \frac{3}{2}^-)$   & $\frac{1}{2}$,$\frac{3}{2}$ &  1    &  1       &   0         &  1   &  0       \\
\label{     }&$                                         $         & $                         $ &       &          &             &      &          \\

\label{D0102}&$\Omega_{c1}(\frac{1}{2}^+,\frac{3}{2}^+)$          & $\frac{1}{2}$,$\frac{3}{2}$ &  1    &  0       &   2         &  2   &  1       \\
\label{D0304}&$\Omega_{c2}(\frac{3}{2}^+,\frac{5}{2}^+)$          & $\frac{3}{2}$,$\frac{5}{2}$ &  2    &  0       &   2         &  2   &  1       \\
\label{D0506}&$\Omega_{c3}(\frac{5}{2}^+,\frac{7}{2}^+)$          & $\frac{5}{2}$,$\frac{7}{2}$ &  3    &  0       &   2         &  2   &  1       \\
\label{D0708}&$\hat\Omega_{c1}(\frac{1}{2}^+,\frac{3}{2}^+)$      & $\frac{1}{2}$,$\frac{3}{2}$ &  1    &  2       &   0         &  2   &  1       \\
\label{D0910}&$\hat\Omega_{c2}(\frac{3}{2}^+,\frac{5}{2}^+)$      & $\frac{3}{2}$,$\frac{5}{2}$ &  2    &  2       &   0         &  2   &  1       \\
\label{D1112}&$\hat\Omega_{c3}(\frac{5}{2}^+,\frac{7}{2}^+)$      & $\frac{5}{2}$,$\frac{7}{2}$ &  3    &  2       &   0         &  2   &  1       \\
\label{D13  }&$\check\Omega_{c0}^{0}(\frac{1}{2}^+)$              & $\frac{1}{2}$               &  0    &  1       &   1         &  0   &  0       \\
\label{D1415}&$\check\Omega_{c1}^{1}(\frac{1}{2}^+,\frac{3}{2}^+)$& $\frac{1}{2}$,$\frac{3}{2}$ &  1    &  1       &   1         &  1   &  0       \\
\label{D1617}&$\check\Omega_{c2}^{1}(\frac{3}{2}^+,\frac{5}{2}^+)$& $\frac{3}{2}$,$\frac{5}{2}$ &  1    &  1       &   1         &  1   &  0       \\
   \hline\hline
\end{tabular}
\label{quantum numbers}
\end{table}

\section{Numerical results \label{Sec: Numerical results}}
In the $^3P_0$ model, $u\bar u$, $d\bar d$ and $s\bar s$ could be created from the vacuum. However, there exists no experimental signal for the decay mode with a $s\bar s$ creation. Once the measured masses of $\Omega_c$ baryons and the mass threshold of the final particles, $\Xi_c^+K^-$ and $\Xi_c^{0}K^0$, have been taken into account, there are two decay channels for $\Omega_c(3000)^0$, $\Omega_c(3050)^0$ and $\Omega_c(3066)^0$. For $\Omega_c(3090)^0$ and $\Omega_c(3119)^0$, $\Xi_c^{'+}K^-$ and $\Xi_c^{'0}K^0 $ are also allowed. Possible decay modes and corresponding hadronic decay widths of these $\Omega_c$ baryons in different $P-$wave and $D-$wave assignments have been computed and presented in from Table~\ref{3000} to Table~\ref{3119}. The vanish modes in these five tables indicate forbidden channels.

\begin{table}[t]
\caption{Decay widths (MeV) of $\Omega_c(3000)^0$ in different assignments.}
\begin{tabular}{p{0.0cm} p{2.5cm}*{3}{p{1.8cm}}}
   \hline\hline
             & $\Omega_{c}(J^P)$                   & $\Xi_c^+K^- $        & $\Xi_c^{0}K^0 $       & $total$            \\
   \hline
\label{P01}  &$\Omega_{c0}(\frac{1}{2}^-)$         & $7.2\times10^{2}$    &  $6.6\times10^{2}$    &  $1.4\times10^{2}$\\
\label{P02}  &$\Omega_{c1}(\frac{1}{2}^-)$         & 0.0                  &  0.0                  &  0.0               \\
\label{P03}  &$\Omega_{c1}(\frac{3}{2}^-)$         & 0.0                  &  0.0                  &  0.0               \\
\label{P04}  &$\Omega_{c2}(\frac{3}{2}^-)$         & 0.3                  &  0.2                  &  0.5               \\
\label{P05}  &$\Omega_{c2}(\frac{5}{2}^-)$         & 0.3                  &  0.2                  &  0.5               \\
\label{P06}  &$\tilde\Omega_{c1}(\frac{1}{2}^-)$   & 0.0                  &  0.0                  &  0.0               \\
\label{P07}  &$\tilde\Omega_{c1}(\frac{3}{2}^-)$   & 0.0                  &  0.0                  &  0.0               \\
\label{   }  &$                                $   &                      &                       &                    \\

\label{D01}  &$\Omega_{c1}(\frac{1}{2}^+)$         & 2.8                  &  2.2                  &  5.0               \\
\label{D02}  &$\Omega_{c1}(\frac{3}{2}^+)$         & 2.8                  &  2.2                  &  5.0               \\
\label{D03}  &$\Omega_{c2}(\frac{3}{2}^+)$         & 0.0                  &  0.0                  &  0.0               \\
\label{D04}  &$\Omega_{c2}(\frac{5}{2}^+)$         & 0.0                  &  0.0                  &  0.0               \\
\label{D05}  &$\Omega_{c3}(\frac{5}{2}^+)$         & $6.9\times10^{-4}$   &  $6.9\times10^{-4}$   &  $1.0\times10^{-3}$\\
\label{D06}  &$\Omega_{c3}(\frac{7}{2}^+)$         & $6.9\times10^{-4}$   &  $6.9\times10^{-4}$   &  $1.0\times10^{-3}$\\
\label{D07}  &$\hat\Omega_{c1}(\frac{1}{2}^+)$     & 25.7                 &  19.4                 &  45.1              \\
\label{D08}  &$\hat\Omega_{c1}(\frac{3}{2}^+)$     & 25.7                 &  19.4                 &  45.1              \\
\label{D09}  &$\hat\Omega_{c2}(\frac{3}{2}^+)$     & 0.0                  &  0.0                  &  0.0               \\
\label{D10}  &$\hat\Omega_{c2}(\frac{5}{2}^+)$     & 0.0                  &  0.0                  &  0.0               \\
\label{D11}  &$\hat\Omega_{c3}(\frac{5}{2}^+)$     & $6.8\times10^{-3}$   &  $3.5\times10^{-3}$   &  $1.3\times10^{-2}$\\
\label{D12}  &$\hat\Omega_{c3}(\frac{7}{2}^+)$     & $6.8\times10^{-3}$   &  $3.5\times10^{-3}$   &  $1.3\times10^{-2}$\\
\label{D13}&$\check\Omega_{c0}^{0}(\frac{1}{2}^+)$ & 0.0                  &  0.0                  &  0.0               \\
\label{D14}&$\check\Omega_{c1}^{1}(\frac{1}{2}^+)$ & 0.0                  &  0.0                  &  0.0               \\
\label{D15}&$\check\Omega_{c1}^{1}(\frac{3}{2}^+)$ & 0.0                  &  0.0                  &  0.0               \\
\label{D16}&$\check\Omega_{c2}^{1}(\frac{3}{2}^+)$ & 0.0                  &  0.0                  &  0.0               \\
\label{D17}&$\check\Omega_{c2}^{1}(\frac{5}{2}^+)$ & 0.0                  &  0.0                  &  0.0               \\

   \hline\hline
\end{tabular}
\label{3000}
\end{table}

\begin{table}[t]
\caption{Decay widths (MeV) of $\Omega_c(3050)^0$ in different assignments.}
\begin{tabular}{p{0.0cm} p{2.5cm}*{3}{p{1.8cm}}}
   \hline\hline
             & $\Omega_{c}(J^P)$                   & $\Xi_c^+K^- $        & $\Xi_c^{0}K^0  $      & $total$            \\
   \hline
\label{P01}  &$\Omega_{c0}(\frac{1}{2}^-)$         & $1.0\times10^{3}$    &  $1.0\times10^{3}$    &  $2.0\times10^{3}$\\
\label{P02}  &$\Omega_{c1}(\frac{1}{2}^-)$         & 0.0                  &  0.0                  &  0.0               \\
\label{P03}  &$\Omega_{c1}(\frac{3}{2}^-)$         & 0.0                  &  0.0                  &  0.0               \\
\label{P04}  &$\Omega_{c2}(\frac{3}{2}^-)$         & 2.9                  &  2.4                  &  5.3               \\
\label{P05}  &$\Omega_{c2}(\frac{5}{2}^-)$         & 2.9                  &  2.4                  &  5.3               \\
\label{P06}  &$\tilde\Omega_{c1}(\frac{1}{2}^-)$   & 0.0                  &  0.0                  &  0.0               \\
\label{P07}  &$\tilde\Omega_{c1}(\frac{3}{2}^-)$   & 0.0                  &  0.0                  &  0.0               \\
\label{   }  &$                                $   &                      &                       &                    \\

\label{D01}  &$\Omega_{c1}(\frac{1}{2}^+)$         & 10.0                 &  8.9                  &  18.9              \\
\label{D02}  &$\Omega_{c1}(\frac{3}{2}^+)$         & 10.0                 &  8.9                  &  18.9              \\
\label{D03}  &$\Omega_{c2}(\frac{3}{2}^+)$         & 0.0                  &  0.0                  &  0.0               \\
\label{D04}  &$\Omega_{c2}(\frac{5}{2}^+)$         & 0.0                  &  0.0                  &  0.0               \\
\label{D05}  &$\Omega_{c3}(\frac{5}{2}^+)$         & $1.4\times10^{-2}$   &  $1.1\times10^{-2}$   &  $2.5\times10^{-2}$\\
\label{D06}  &$\Omega_{c3}(\frac{7}{2}^+)$         & $1.4\times10^{-2}$   &  $1.1\times10^{-2}$   &  $2.5\times10^{-2}$\\
\label{D07}  &$\hat\Omega_{c1}(\frac{1}{2}^+)$     & 89.4                 &  80.4                 &  168.8             \\
\label{D08}  &$\hat\Omega_{c1}(\frac{3}{2}^+)$     & 89.4                 &  80.4                 &  168.8             \\
\label{D09}  &$\hat\Omega_{c2}(\frac{3}{2}^+)$     & 0.0                  &  0.0                  &  0.0               \\
\label{D10}  &$\hat\Omega_{c2}(\frac{5}{2}^+)$     & 0.0                  &  0.0                  &  0.0               \\
\label{D11}  &$\hat\Omega_{c3}(\frac{5}{2}^+)$     & 0.1                  &  0.1                  &  0.2 \\
\label{D12}  &$\hat\Omega_{c3}(\frac{7}{2}^+)$     & 0.1                  &  0.1                  &  0.2 \\
\label{D13}&$\check\Omega_{c0}^{0}(\frac{1}{2}^+)$ & 0.0                  &  0.0                  &  0.0               \\
\label{D14}&$\check\Omega_{c1}^{1}(\frac{1}{2}^+)$ & 0.0                  &  0.0                  &  0.0               \\
\label{D15}&$\check\Omega_{c1}^{1}(\frac{3}{2}^+)$ & 0.0                  &  0.0                  &  0.0               \\
\label{D16}&$\check\Omega_{c2}^{1}(\frac{3}{2}^+)$ & 0.0                  &  0.0                  &  0.0               \\
\label{D17}&$\check\Omega_{c2}^{1}(\frac{5}{2}^+)$ & 0.0                  &  0.0                  &  0.0               \\

   \hline\hline
\end{tabular}
\label{3050}
\end{table}

\begin{table}[t]
\caption{Decay widths (MeV) of $\Omega_c(3066)^0$ in different assignments.}
\begin{tabular}{p{0.0cm} p{2.5cm}*{3}{p{1.8cm}}}
   \hline\hline
             & $\Omega_{c}(J^P)$                   & $\Xi_c^+K^- $        & $\Xi_c^{0}K^0  $      & $total$           \\
   \hline
\label{P01}  &$\Omega_{c0}(\frac{1}{2}^-)$         & $1.1\times10^{3}$    &  $1.1\times10^{3}$    &  $2.2\times10^{3}$ \\
\label{P02}  &$\Omega_{c1}(\frac{1}{2}^-)$         & 0.0                  &  0.0                  &  0.0               \\
\label{P03}  &$\Omega_{c1}(\frac{3}{2}^-)$         & 0.0                  &  0.0                  &  0.0               \\
\label{P04}  &$\Omega_{c2}(\frac{3}{2}^-)$         & 4.6                  &  3.9                  &  8.5               \\
\label{P05}  &$\Omega_{c2}(\frac{5}{2}^-)$         & 4.6                  &  3.9                  &  8.5               \\
\label{P06}  &$\tilde\Omega_{c1}(\frac{1}{2}^-)$   & 0.0                  &  0.0                  &  0.0               \\
\label{P07}  &$\tilde\Omega_{c1}(\frac{3}{2}^-)$   & 0.0                  &  0.0                  &  0.0               \\
\label{   }  &$                                $   &                      &                       &                    \\

\label{D01}  &$\Omega_{c1}(\frac{1}{2}^+)$         & 12.8                 &  11.7                 &  24.5              \\
\label{D02}  &$\Omega_{c1}(\frac{3}{2}^+)$         & 12.8                 &  11.7                 &  24.5              \\
\label{D03}  &$\Omega_{c2}(\frac{3}{2}^+)$         & 0.0                  &  0.0                  &  0.0               \\
\label{D04}  &$\Omega_{c2}(\frac{5}{2}^+)$         & 0.0                  &  0.0                  &  0.0               \\
\label{D05}  &$\Omega_{c3}(\frac{5}{2}^+)$         & $2.6\times10^{-2}$   &  $2.1\times10^{-2}$   &  $4.7\times10^{-2}$\\
\label{D06}  &$\Omega_{c3}(\frac{7}{2}^+)$         & $2.6\times10^{-2}$   &  $2.1\times10^{-2}$   &  $4.7\times10^{-2}$\\
\label{D07}  &$\hat\Omega_{c1}(\frac{1}{2}^+)$     & $1.1\times10^{2}$    &  $1.0\times10^{2}$    &  $2.1\times10^{2}$ \\
\label{D08}  &$\hat\Omega_{c1}(\frac{3}{2}^+)$     & $1.1\times10^{2}$    &  $1.0\times10^{2}$    &  $2.1\times10^{2}$ \\
\label{D09}  &$\hat\Omega_{c2}(\frac{3}{2}^+)$     & 0.0                  &  0.0                  &  0.0               \\
\label{D10}  &$\hat\Omega_{c2}(\frac{5}{2}^+)$     & 0.0                  &  0.0                  &  0.0               \\
\label{D11}  &$\hat\Omega_{c3}(\frac{5}{2}^+)$     & 0.3                  &  0.2                  &  0.5               \\
\label{D12}  &$\hat\Omega_{c3}(\frac{7}{2}^+)$     & 0.3                  &  0.2                  &  0.5               \\
\label{D13}&$\check\Omega_{c0}^{0}(\frac{1}{2}^+)$ & 0.0                  &  0.0                  &  0.0               \\
\label{D14}&$\check\Omega_{c1}^{1}(\frac{1}{2}^+)$ & 0.0                  &  0.0                  &  0.0               \\
\label{D15}&$\check\Omega_{c1}^{1}(\frac{3}{2}^+)$ & 0.0                  &  0.0                  &  0.0               \\
\label{D16}&$\check\Omega_{c2}^{1}(\frac{3}{2}^+)$ & 0.0                  &  0.0                  &  0.0               \\
\label{D17}&$\check\Omega_{c2}^{1}(\frac{5}{2}^+)$ & 0.0                  &  0.0                  &  0.0               \\

   \hline\hline
\end{tabular}
\label{3066}
\end{table}

\begin{table}[t]
\caption{Decay widths (MeV) of $\Omega_c(3090)^0$ in different assignments.}
\begin{tabular}{p{-0.5cm} p{1.2cm}*{5}{p{1.35cm}}}
   \hline\hline
             & $\Omega_{c}(J^P)$                   & $\Xi_c^+K^- $        & $\Xi_c^{0}K^0 $     & $\Xi_c^{'+}K^- $   & $\Xi_c^{'0}K^0 $   & $total$            \\
   \hline
\label{P01}  &$\Omega_{c0}(\frac{1}{2}^-)$         & $1.2\times10^{3}$    &  $1.1\times10^{3}$  &  0.0               &  0.0               &  $2.3\times10^{3}$ \\
\label{P02}  &$\Omega_{c1}(\frac{1}{2}^-)$         & 0.0                  &  0.0                &  $3.6\times10^{2}$ &  $3.1\times10^{2}$ &  $6.7\times10^{2}$ \\
\label{P03}  &$\Omega_{c1}(\frac{3}{2}^-)$         & 0.0                  &  0.0                &  $3.0\times10^{-2}$&  $1.2\times10^{-2}$&  $4.2\times10^{-2}$\\
\label{P04}  &$\Omega_{c2}(\frac{3}{2}^-)$         & 8.0                  &  7.0                &  $5.3\times10^{-2}$&  $2.2\times10^{-2}$&  15.1              \\
\label{P05}  &$\Omega_{c2}(\frac{5}{2}^-)$         & 8.0                  &  7.0                &  $2.3\times10^{-2}$&  $9.9\times10^{-3}$&  15.1              \\
\label{P06}  &$\tilde\Omega_{c1}(\frac{1}{2}^-)$   & 0.0                  &  0.0                &  $5.4\times10^{2}$ &  $4.6\times10^{2}$ &  $1.0\times10^{3}$ \\
\label{P07}  &$\tilde\Omega_{c1}(\frac{3}{2}^-)$   & 0.0                  &  0.0                &  0.2               &  0.1               &  0.3               \\
\label{   }  &$                                $   &                      &                     &                    &                    &                    \\

\label{D01}  &$\Omega_{c1}(\frac{1}{2}^+)$         & 17.4                 &  16.3               &  0.4               &  0.2               &  34.3              \\
\label{D02}  &$\Omega_{c1}(\frac{3}{2}^+)$         & 17.4                 &  16.3               &  0.1               &  0.1               &  33.9              \\
\label{D03}  &$\Omega_{c2}(\frac{3}{2}^+)$         & 0.0                  &  0.0                &  0.9               &  0.5               &  1.4               \\
\label{D04}  &$\Omega_{c2}(\frac{5}{2}^+)$         & 0.0                  &  0.0                &  $4.0\times10^{-5}$&  $1.1\times10^{-5}$&  $5.1\times10^{-5}$\\
\label{D05}  &$\Omega_{c3}(\frac{5}{2}^+)$         & $5.7\times10^{-2}$   &  $4.8\times10^{-2}$ &  $4.5\times10^{-5}$&  $1.3\times10^{-5}$&  0.1               \\
\label{D06}  &$\Omega_{c3}(\frac{7}{2}^+)$         & $5.7\times10^{-2}$   &  $4.8\times10^{-2}$ &  $2.5\times10^{-5}$&  $7.6\times10^{-6}$&  0.1               \\
\label{D07}  &$\hat\Omega_{c1}(\frac{1}{2}^+)$     & $1.6\times10^{2}$    &  $1.5\times10^{2}$  &  3.4               &  2.0               &  $3.1\times10^{2}$ \\
\label{D08}  &$\hat\Omega_{c1}(\frac{3}{2}^+)$     & $1.6\times10^{2}$    &  $1.5\times10^{2}$  &  0.8               &  0.5               &  $3.1\times10^{2}$ \\
\label{D09}  &$\hat\Omega_{c2}(\frac{3}{2}^+)$     & 0.0                  &  0.0                &  7.6               &  4.6               &  12.6              \\
\label{D10}  &$\hat\Omega_{c2}(\frac{5}{2}^+)$     & 0.0                  &  0.0                &  $3.9\times10^{-4}$&  $1.2\times10^{-4}$&  $5.1\times10^{-2}$\\
\label{D11}  &$\hat\Omega_{c3}(\frac{5}{2}^+)$     & 0.6                  &  0.5                &  $4.5\times10^{-4}$&  $1.3\times10^{-4}$&  1.0               \\
\label{D12}  &$\hat\Omega_{c3}(\frac{7}{2}^+)$     & 0.6                  &  0.5                &  $2.5\times10^{-4}$&  $7.5\times10^{-5}$&  1.0               \\
\label{D13}&$\check\Omega_{c0}^{0}(\frac{1}{2}^+)$ & 0.0                  &  0.0                &  6.7               &  4.1               &  10.8              \\
\label{D14}&$\check\Omega_{c1}^{1}(\frac{1}{2}^+)$ & 0.0                  &  0.0                &  0.0               &  0.0               &  0.0               \\
\label{D15}&$\check\Omega_{c1}^{1}(\frac{3}{2}^+)$ & 0.0                  &  0.0                &  0.0               &  0.0               &  0.0               \\
\label{D16}&$\check\Omega_{c2}^{1}(\frac{3}{2}^+)$ & 0.0                  &  0.0                &  3.4               &  2.0               &  5.4               \\
\label{D17}&$\check\Omega_{c2}^{1}(\frac{5}{2}^+)$ & 0.0                  &  0.0                &  $3.9\times10^{-4}$&  $1.2\times10^{-4}$&  $5.1\times10^{-4}$\\

   \hline\hline
\end{tabular}
\label{3090}
\end{table}

\begin{table}[t]
\caption{Decay widths (MeV) of $\Omega_c(3119)^0$ in different assignments.}
\begin{tabular}{p{-0.5cm} p{1.2cm}*{5}{p{1.35cm}}}
   \hline\hline
             & $\Omega_{c}(J^P)$                   & $\Xi_c^+K^- $        & $\Xi_c^{0}K^0 $     & $\Xi_c^{'+}K^- $   & $\Xi_c^{'0}K^0 $   & $total$            \\
   \hline
\label{P01}  &$\Omega_{c0}(\frac{1}{2}^-)$         & $1.2\times10^{3}$    &  $1.2\times10^{3}$  &  0.0               &  0.0               &  $2.4\times10^{3}$ \\
\label{P02}  &$\Omega_{c1}(\frac{1}{2}^-)$         & 0.0                  &  0.0                &  $5.4\times10^{2}$ &  $5.1\times10^{2}$ &  $10.5\times10^{2}$\\
\label{P03}  &$\Omega_{c1}(\frac{3}{2}^-)$         & 0.0                  &  0.0                &  0.2               &  0.2               &  0.5               \\
\label{P04}  &$\Omega_{c2}(\frac{3}{2}^-)$         & 13.9                 &  12.5               &  0.4               &  0.4               &  27.3              \\
\label{P05}  &$\Omega_{c2}(\frac{5}{2}^-)$         & 13.9                 &  12.5               &  0.2               &  0.2               &  26.8              \\
\label{P06}  &$\tilde\Omega_{c1}(\frac{1}{2}^-)$   & 0.0                  &  0.0                &  $8.1\times10^{2}$ &  $7.7\times10^{2}$ &  $1.6\times10^{3}$ \\
\label{P07}  &$\tilde\Omega_{c1}(\frac{3}{2}^-)$   & 0.0                  &  0.0                &  1.6               &  1.2               &  2.8               \\
\label{   }  &$                                $   &                      &                     &                    &                    &                    \\

\label{D01}  &$\Omega_{c1}(\frac{1}{2}^+)$         & 23.5                 &  22.2               &  1.4               &  1.2               &  48.3              \\
\label{D02}  &$\Omega_{c1}(\frac{3}{2}^+)$         & 23.4                 &  22.2               &  0.4               &  0.3               &  46.3              \\
\label{D03}  &$\Omega_{c2}(\frac{3}{2}^+)$         & 0.0                  &  0.0                &  3.2               &  2.6               &  5.8               \\
\label{D04}  &$\Omega_{c2}(\frac{5}{2}^+)$         & 0.0                  &  0.0                &  $9.0\times10^{-4}$&  $5.8\times10^{-4}$&  $1.5\times10^{-3}$\\
\label{D05}  &$\Omega_{c3}(\frac{5}{2}^+)$         & 0.1                  &  0.1                &  $1.0\times10^{-3}$&  $6.6\times10^{-4}$&  0.2               \\
\label{D06}  &$\Omega_{c3}(\frac{7}{2}^+)$         & 0.1                  &  0.1                &  $5.8\times10^{-4}$&  $3.7\times10^{-4}$&  0.2               \\
\label{D07}  &$\hat\Omega_{c1}(\frac{1}{2}^+)$     & $2.1\times10^{2}$    &  $2.0\times10^{2}$  &  12.6              &  10.6              &  $4.3\times10^{2}$ \\
\label{D08}  &$\hat\Omega_{c1}(\frac{3}{2}^+)$     & $2.1\times10^{2}$    &  $2.0\times10^{2}$  &  3.2               &  2.6               &  $4.1\times10^{2}$ \\
\label{D09}  &$\hat\Omega_{c2}(\frac{3}{2}^+)$     & 0.0                  &  0.0                &  28.5              &  23.7              &  52.3              \\
\label{D10}  &$\hat\Omega_{c2}(\frac{5}{2}^+)$     & 0.0                  &  0.0                &  $8.9\times10^{-3}$&  $5.7\times10^{-3}$&  $1.4\times10^{-2}$\\
\label{D11}  &$\hat\Omega_{c3}(\frac{5}{2}^+)$     & 1.2                  &  1.1                &  $1.0\times10^{-2}$&  $6.5\times10^{-3}$&  2.3               \\
\label{D12}  &$\hat\Omega_{c3}(\frac{7}{2}^+)$     & 1.2                  &  1.1                &  $5.8\times10^{-3}$&  $3.7\times10^{-3}$&  2.3               \\
\label{D13}&$\check\Omega_{c0}^{0}(\frac{1}{2}^+)$ & 0.0                  &  0.0                &  25.2              &  21.0              &  46.2              \\
\label{D14}&$\check\Omega_{c1}^{1}(\frac{1}{2}^+)$ & 0.0                  &  0.0                &  0.0               &  0.0               &  0.0               \\
\label{D15}&$\check\Omega_{c1}^{1}(\frac{3}{2}^+)$ & 0.0                  &  0.0                &  0.0               &  0.0               &  0.0               \\
\label{D16}&$\check\Omega_{c2}^{1}(\frac{3}{2}^+)$ & 0.0                  &  0.0                &  12.7              &  10.6              &  23.2              \\
\label{D17}&$\check\Omega_{c2}^{1}(\frac{5}{2}^+)$ & 0.0                  &  0.0                &  $9.0\times10^{-3}$&  $5.7\times10^{-3}$&  $1.5\times10^{-2}$\\

   \hline\hline
\end{tabular}
\label{3119}
\end{table}

All the five new $\Omega_c^0$ baryons were observed in the $\Xi_c^+K^-$ channel. The decay width of $\Omega_c(3000)^0$ is $\Gamma_{\Omega_{c}(3000)^{0}}=(4.5 \pm 0.6 \pm 0.3)$ MeV. From the results in Table~\ref{3000}, the most possible assignment of $\Omega_c(3000)^0$ is $\Omega_{c1}(\frac{1}{2}^+)$ or $\Omega_{c1}(\frac{3}{2}^+)$. In these two assignments, the total decay widths are the same $5.0$ MeV, which is very close to the experimental data.

The experimental measured decay width of $\Omega_c(3050)^0$ is extremely narrow, which is smaller than $1.0$ MeV. From Table~\ref{3050}, the possible assignment of $\Omega_c(3050)^0$ is $\hat\Omega_{c3}(\frac{5}{2}^+)$ or $\hat\Omega_{c3}(\frac{7}{2}^+)$. The theoretical prediction of the decay width is $0.2$ MeV.

For $\Omega_c(3066)^0$, the measured decay width is $(3.5 \pm 0.4 \pm 0.2)~{\rm MeV}$. From Table~\ref{3066}, the total decay widths of $\Omega_{c2}(\frac{3}{2}^-)$ and $\Omega_{c2}(\frac{5}{2}^-)$ are the same $8.5$ MeV, which is a little larger than the experimental result. Under the theoretical uncertainty, these assignments are possible for $\Omega_c(3066)^0$.

For $\Omega_c(3090)^0$ and $\Omega_c(3119)^0$, the channels $\Xi_c^{'+}K^-$ and $\Xi_c^{'0}K^0$ open. In Table~\ref{3090} and Table~\ref{3119}, many channels decaying to $\Xi_c^+K^-$ and $\Xi_c^0K^0$ are forbidden while to $\Xi_c^{'+}K^-$ and $\Xi_c^{'0}K^0$ are not. In particular, the total decay widths of $\Omega_c(3090)^0$ and $\Omega_c(3119)^0$ in some assignments are extremely large. In comparison with experiment, the possible assignment for $\Omega_c(3090)^0$ is $\Omega_{c2}(\frac{3}{2}^-)$ or $\Omega_{c2}(\frac{5}{2}^-)$, and four $D-$wave assignments are possible for $\Omega_c(3119)^0$: $\Omega_{c3}(\frac{5}{2}^+)$, $\Omega_{c3}(\frac{7}{2}^+)$, $\hat\Omega_{c3}(\frac{5}{2}^+)$ and $\hat\Omega_{c3}(\frac{7}{2}^+)$. In these assignments, the decay widths to $\Xi_c^{'+}K^-$ and $\Xi_c^{'0}K^0$ channels are very tiny and could be neglected.

\section{Conclusions and discussions\label{Sec: summary}}
In this work, we study the hadronic decays of five newly observed baryons $\Omega_c(3000)^0$, $\Omega_c(3050)^0$, $\Omega_c(3066)^0$, $\Omega_c(3090)^0$ and $\Omega_c(3119)^0$ in the $^3P_0$ model. In different assignments of $1P-$wave and $1D-$wave $\Omega_c$, hadronic decay widths of these baryons have been calculated. In comparison with experiment, possible assignments of these $\Omega_c^0$ have been made.

All the five $\Omega_c^0$ baryons were observed in the $\Xi_c^+K^-$ channel with very narrow decay widths. Our results indicates that these $\Omega_c^0$ could be $P-$wave or $D-$wave $\Omega_c^0$ baryons.
$\Omega_c(3000)^0$ is possibly $1D-$wave $\Omega_{c1}(\frac{1}{2}^+)$ or $\Omega_{c1}(\frac{3}{2}^+)$. $\Omega_c(3050)^0$ is possibly $1D-$wave $\hat\Omega_{c3}(\frac{5}{2}^+)$ or $\hat\Omega_{c3}(\frac{7}{2}^+)$. $\Omega_c(3066)^0$ is possibly $1P-$wave $\Omega_{c2}(\frac{3}{2}^-)$ or $\Omega_{c2}(\frac{5}{2}^-)$. The possible assignment for $\Omega_c(3090)^0$ is $1P-$wave $\Omega_{c2}(\frac{3}{2}^-)$ or $\Omega_{c2}(\frac{5}{2}^-)$, and four $D-$wave assignments $\Omega_{c3}(\frac{5}{2}^+)$, $\Omega_{c3}(\frac{7}{2}^+)$, $\hat\Omega_{c3}(\frac{5}{2}^+)$ and $\hat\Omega_{c3}(\frac{7}{2}^+)$ are possible for $\Omega_c(3119)^0$. The predicted decay widths are consistent with experimental data.

In experiment, only $\Omega_c^0 \to \Xi_c^+K^-$ were observed, which may not provide enough information on their identification. Furthermore, resonance with the same $J^P$ numbers and similar masses may mix with each other, which may make it difficult to distinguish them. As for $1D-$wave $\Omega_c$, the channels $\Xi_c^{'+}K^-$ and $\Xi_c^{'0}K^0$ is hardly to be observed for the tiny decay widths.  More information on these $\Omega_c$ baryons are expected to be given in forthcoming experiment.

\begin{acknowledgments}
This work is supported by National Natural Science Foundation of China under the grants: 11475111 and 11075102.
\end{acknowledgments}

\end{document}